\documentclass{agnspec}
\usepackage{graphics}
\usepackage{psfig}

\def\NH{$N_{\rm H}$}
\def\LX{$L_{\rm X}$}
\def\LHa{$L_{{\rm H} \alpha}$}

\def\eps{ergs s$^{-1}$}
\def\pcm{cm$^{-2}$}

\def\chandra{\it Chandra}
\def\asca{\it ASCA}

\def\HST{\it HST}

\def\gtsima{$\; \buildrel > \over \sim \;$}
\def\simgt{\lower.5ex\hbox{\gtsima}}
\def\ltsima{$\; \buildrel < \over \sim \;$}
\def\simlt{\lower.5ex\hbox{\ltsima}}

\begin{document}
\title{Chandra Snapshot Observations of LINERs with a Compact Radio Core}
\author{Y. Terashima\inst{1,2} \and A.S. Wilson\inst{2,3}}
\institute{
Institute of Space and Astronautical Science, 3-1-1 Yoshinodai,
 Sagamihara, Kanagawa 229-8510, Japan
\and
Astronomy Department, University of Maryland, College Park, MD 20742, USA
\and
Space Telescope Science Institute, 3700 San Martin Drive, Baltimore, MD 21218, 
USA
}
\maketitle

\begin{abstract}

The results of {\chandra} snapshot observations of 11 LINERs
(Low-Ionization Nuclear Emission-line Regions), three low-luminosity
Seyfert galaxies, and one HII-LINER transition object are
presented. Our sample consists of all the objects with a flat or
inverted spectrum compact radio core in the VLA survey of 48
low-luminosity AGN (LLAGN) by Nagar et al. (2000). An X-ray nucleus is
detected in all galaxies except one and their X-ray luminosities are
in the range $5\times10^{38}$ to $8\times10^{41}$ {\eps}. The X-ray to
H$\alpha$ luminosity ratios for 11 out of 14 objects are in good
agreement with the value characteristic of LLAGNs and more luminous
AGNs, and indicate that their optical emission lines are predominantly
powered by a LLAGN. For three objects, this ratio is less than
expected.  Comparing with multi-wavelength results, we find that these
three galaxies are most likely to be heavily obscured AGN. We compare
the radio to X-ray luminosity ratio of LLAGNs with those of
more-luminous AGNs, and confirm the suggestion that a large fraction
of LLAGNs are radio loud.

\textbf{}

\end{abstract}

\section{Introduction}

Low-Ionization nuclear emission-line regions (LINERs) are found in
many nearby bright galaxies (e.g., Ho, Filippenko, \& Sargent
1997a). Extensive studies in various wavelengths have shown that type
1 LINERs (LINER 1s, i.e., those galaxies having broad H$\alpha$ and
possibly other broad Balmer lines in their nuclear optical spectra)
are powered by a low-luminosity AGN (LLAGN) with a bolometric
luminosity less than $\sim10^{42}$ {\eps} (Ho et al. 2001; Terashima,
Ho, \& Ptak 2000a; Ho et al. 1997b). On the other hand, the energy
source of LINER 2s is likely to be heterogeneous.  Some LINER 2s show
clear signatures of the presence of an AGN, while others are most
probably powered by stellar processes, and the luminosity ratio
{\LX}/{\LHa} can be used to discriminate between these power sources
(e.g., Terashima et al. 2000b).  It is interesting
to note that currently there are only a few LINER 2s known to host an
obscured AGN (e.g., Turner et al. 2001). This paucity of obscured AGN
in LINERs may indicate that LINER 2s are not simply a low-luminosity
extension of luminous Seyfert 2s, which often show heavy obscuration
with a column density averaging {\NH} $\sim$ $10^{23}$ {\pcm} (e.g.,
Turner et al. 1997). Alternatively, biases against finding heavily
obscured LLAGNs may be important. For example, objects selected
through optical emission lines or X-ray fluxes are probably biased in
favor of less absorbed ones, even if one uses the X-ray band above 2
keV.

  In contrast, radio observations, particularly at high frequency, are
much less affected by absorption. Nagar et al. (2002) have reported a
VLA 2 cm radio survey of all 96 LLAGNs within a distance of 19
Mpc. These LLAGNs come from the Palomar spectroscopic survey of bright
galaxies (Ho et al. 1997a). As a pilot study of the X-ray properties
of LLAGNs, we report here a {\chandra} survey of a subset, comprising
14 galaxies, of Nagar et al's (2002) sample. 
We have detected 13 of the
galactic nuclei with {\chandra}. We also examine the ``radio
loudness'' of our sample and compare it with other classes of AGN.

\section{The Sample and Observations}

Our sample is based on the 15~GHz VLA observations by Nagar et
al. (2000). Their sample of 48 objects consists of 22 LINERs, 18
transition objects between LINERs and HII nuclei, and eight
low-luminosity Seyferts selected from the optical spectroscopic survey
of Ho et al. (1997a). 

  We selected 14 objects showing a flat to inverted spectrum radio
core ($\alpha \ge -0.3$, $S_{\nu}\propto \nu^{\alpha}$) according to
Nagar et al.'s (2000) comparison with longer wavelength radio data
published in literature. The targets are summarized in Table 1. The
sample consists of seven LINER 1s, three LINER 2s, two Seyfert 1s, one
Seyfert 2, and one transition 2 object. 12 out of these 14 objects
have been observed with the VLBA and high brightness temperature
($T_{\rm b}>10^7$ K) radio cores were detected in all of them (Falcke
et al. 2000; Ulvestad \& Ho 2001; Nagar et al. 2002).

The exposure time was typically two ksec each. All the objects were
observed with the ACIS-S3 back-illuminated CCD
chip. Eight objects were observed in 1/8 sub-frame mode (frame time
0.4 s) to minimize effects of pileup. 1/2 sub-frame modes were used
for three objects. Detailed results are given in Terashima \& Wilson
(2002c).

\begin{table}
      \caption{The Sample.}
	\begin{tabular}{ll}
            \hline
            \noalign{\smallskip}
            Class	&  Name\\
            \noalign{\smallskip}
            \hline
            \noalign{\smallskip}
            LINER 1             & NGC 266, 2787, 3226, 4143, 4203, 4278, 4579\\
            LINER 2             & NGC 3169, 4548, 6500\\
            Seyfert 1           & NGC 4565, 5033\\
            Seyfert 2           & NGC 3147\\
	    Transition 2	& NGC 5866\\
            \noalign{\smallskip}
             \hline
	\end{tabular}
\end{table}

\section{Results}

An X-ray nucleus is seen in all the galaxies except for NGC 5866.  The
X-ray luminosities corrected for absorption are in the range
$5\times10^{38}$ to $8\times10^{41}$ {\eps}.  The positions of the
X-ray nuclei coincide with the radio core positions to within the
positional accuracy of {\chandra}. 

Spectral fits were performed for relatively bright objects. The
pileup effect for the three objects with the largest count rate per
frame (NGC 4203, NGC 4579, and NGC 5033) is serious and we did not
attempt detailed spectral fits. Instead, we use the spectra and fluxes
measured with {\asca} for these three objects (Terashima et al. 2002b
and references therein) in the following discussions. We confirmed
that the nuclear X-ray source dominates the hard X-ray emission within
the beam size of {\asca}.

A power-law model modified by absorption was applied and acceptable
fits were obtained in all cases. The photon indices of the nuclear
sources are generally consistent with the typical values observed in
LLAGNs (photon index $\Gamma = 1.6-2.0$, e.g., Terashima et al. 2002a,
2002b), although errors are quite large due to the limited photon
statistics.  The two objects (LINER 2s NGC 3169 and NGC 4548) show
large absorption column density {\NH}=$1.1\times10^{23}$ {\pcm} and
$1.6\times10^{22}$ {\pcm} , respectively, while NGC 3226 is less
absorbed ({\NH}=$9.3\times10^{21}$ {\pcm}).  Others have small column
densities which are consistent with `type 1' AGNs.  NGC 2787 has only
8 detected photons in the 0.5--8 keV band and is too faint to obtain
spectral information.

\section{Discussion}

\subsection{Power Source of LINERs}

We test whether the detected X-ray sources are the power source of
their optical emission lines by examining the luminosity ratio
{\LX}/{\LHa}. The H$\alpha$ luminosities ({\LHa}) were taken from Ho
et al. (1997a) and corrected for the reddening estimated from the
Balmer decrement for the narrow lines. The X-ray luminosities ({\LX})
in the $2-10$ keV band, and corrected for absorption, were used.  The
resulting {\LX}/{\LHa} ratios of most objects are in the range of AGNs
($\log$ {\LX}/{\LHa} {\simgt}1) and in good agreement with the strong
correlation between {\LX} and {\LHa} for LLAGNs, luminous Seyferts,
and QSOs presented in Terashima et al. (2000a) and Ho et al. (2001).
This indicates that their optical emission lines are predominantly
powered by a LLAGN.

The three objects NGC 2787, NGC 5866, and NGC 6500, however, have much
lower {\LX}/{\LHa} ratios ($\log$ {\LX}/{\LHa} $\simlt$0) than expected
from the correlation, and their X-ray
luminosities are not enough to power the H$\alpha$ luminosities.
This X-ray faintness could indicate one or
more of several possibilities such as (1) an AGN is the power source,
but is heavily absorbed at energies above 2 keV, (2) an AGN is the
power source, but is currently switched-off or in a faint state, and
(3) the optical narrow emission lines are powered by some other source(s)
than an AGN.


If an AGN is present in these X-ray faint objects and absorbed in the
hard energy band above 2 keV, only scattered and/or highly absorbed
X-rays can be observed, and then the intrinsic luminosity would be
much higher than that observed. This can account for the low
{\LX}/{\LHa} ratios and high radio to X-ray luminosity ratios ($\nu
L_{\nu}$(5 GHz)/{\LX}). If the intrinsic X-ray luminosities are about
one or two orders of magnitude higher than those observed, as is often
inferred for Seyfert 2 galaxies, {\LX}/{\LHa} and $\nu L_{\nu}$(5
GHz)/{\LX} become typical of LLAGNs.

Additional lines of evidence which support the presence of an AGN
include the fact that all three of these galaxies (NGC 2787, NGC 5866, and
NGC 6500) have VLBI-detected, sub-pc scale, nuclear radio core sources
(Falcke et al. 2000), a broad H$\alpha$ component (in NGC 2787, and an
ambiguous detection in NGC 5866; Ho et al.  1997b), a variable radio
core in NGC 2787, and a jet-like linear structure in a high-resolution
radio map at 5 GHz with the VLBA (NGC 6500; Falcke et al. 2000). Only
an upper limit to the X-ray flux is obtained for NGC 5866. If an X-ray
nucleus is present in this galaxy and its luminosity is only slightly
below the upper limit, this source could be an AGN obscured by a
column density {\NH}$\sim10^{23}$ {\pcm} or larger.  If the intrinsic
luminosity of the nucleus is {\it much} lower than the observed upper
limit, an AGN would have to be almost completely obscured and/or the
optical emission lines powered by some other source(s). The optical
classification (transition object) suggests the presence of an
ionizing source other than an AGN.

\begin{figure}
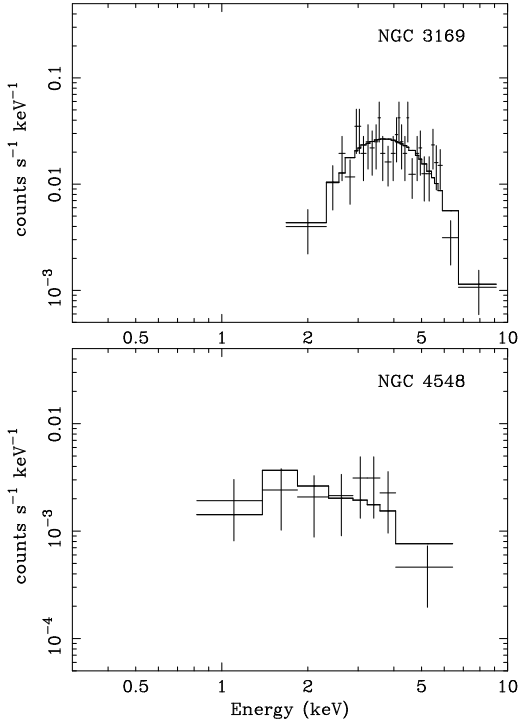

\centerline{\psfig{figure=yuichi_terashima_fig1a.ps,width=6.8cm,angle=-90}}
\centerline{\psfig{figure=yuichi_terashima_fig1b.ps,width=6.8cm,angle=-90}}
\caption[]{Examples of Chandra spectra. (a) NGC 3169 and (b) NGC 4548}
\end{figure}

\subsection{Obscured LLAGNs}

In our sample, we found at least two highly absorbed LLAGNs (NGC 3169
and NGC 4548). In addition, if the X-ray faint objects discussed in
the previous subsection are indeed AGNs, they are most probably highly
absorbed with {\NH}$>10^{23}$ {\pcm}. Among these absorbed objects,
NGC 2787 is classified as a LINER 1.9, NGC 3169, NGC 4548, and NGC
6500 as LINER 2s, and NGC 5866 as a transition 2 object. Thus, heavily
absorbed LINER 1.9s/2s, of which few are known, are found in the present
observations demonstrating that radio selection is a valuable
technique for finding obscured AGNs. Along with heavily obscured
LLAGNs known in low-luminosity Seyfert 2s (e.g., NGC 2273, NGC 2655,
NGC 3079, NGC 4941, and NGC 5194; Terashima et al. 2002a), our
observations show that at least some type 2 LLAGNs are simply
low-luminosity counterparts of luminous Seyferts in which heavy
absorption is often observed. Some LINER 2s (e.g., NGC 4594, Terashima
et al. 2002a; NGC 4374, Finoguenov \& Jones 2001; NGC 4486, Wilson \&
Yang 2002) and low-luminosity Seyfert 2s (NGC 3147) show no strong
absorption.  Therefore, the orientation dependent unification scheme
does not always apply to AGNs in the low-luminosity regime.

\subsection{Radio Loudness of LLAGNs}

Earlier studies have suggested that LLAGNs tend to be radio loud
compared to more luminous AGNs based on the spectral energy
distributions of seven LLAGNs (Ho 1999) and, for a larger sample, on
the conventional definition of radio loudness $R_{\rm O}=L_{\nu}$(5
GHz)/$L_{\nu}$(B) (the subscript ``O'' stands for optical), with
$R_{\rm O}>10$ being radio loud (Ho \& Peng 2001).  Ho \& Peng (2001)
measured the luminosities of the nuclei by spatial analysis of optical
images obtained with {\HST} to reduce the contribution from stellar
light. A caveat in the use of optical measurements for the definition
of radio loudness is extinction, which will lead to an overestimate of
$R_{\rm O}$. Although Ho \& Peng (2001) used only type 1--1.9 objects,
some objects of these types show high absorption columns in their
X-ray spectra. In this subsection, we study radio loudness by
comparing radio and hard X-ray luminosities. Since the unabsorbed
luminosity for objects with {\NH} \simgt $10^{23}$ {\pcm} (equivalent
to $A_{\rm V}$ \simgt 50 mag for a normal gas to dust ratio) can be
reliably measured in the 2--10 keV band, it is clear that replacement
of optical by hard X-ray luminosity potentially yields considerable
advantages.

In the following analysis, radio data at 5 GHz taken from the
literature are used since fluxes at this frequency are widely
available for various classes of objects.  We used the radio
luminosities primarily obtained with the VLA at \simlt
$1^{\prime\prime}$ resolution for the present sample. High resolution
VLA data at 5 GHz are not available for several objects.  For four
objects among such cases, VLBA observations at 5 GHz with 150 mas
resolution are published in the literature (Falcke et al. 2000) and
are used here. For two objects, we estimated 5 GHz fluxes from 15 GHz
data by assuming a spectral slope of $\alpha=0$ (cf. Nagar et
al. 2001). Since our sample is selected based on the presence of a
compact radio core, the sample could be biased to more radio loud
objects. Therefore, we constructed a larger sample by adding objects
taken from the literature for which 5 GHz radio, 2--10 keV X-ray, and
$R_{\rm O}$ measurements are available.

First, we introduce the ratio $R_{\rm X} = \nu L_{\nu}$(5 GHz)/{\LX}
as a measure of radio loudness and compare the ratio with the
conventional $R_{\rm O}$ parameter. The X-ray luminosity {\LX} in the
2--10 keV band (source rest frame), corrected for absorption, is used.
We examine the behavior of $R_{\rm X}$ using samples of AGN over a
wide range of luminosity, including LLAGN, the Seyfert sample of Ho \&
Peng (2001) and PG quasars which are also used in their analysis.
The X-ray luminosities (mostly measured with {\asca}) are
compiled from the literature.

Fig. 2 compares the parameters $R_{\rm O}$ and $R_{\rm X}$ for the
Seyferts and PG sample. These two parameters correlate well for most
Seyferts.  Some Seyferts have higher $R_{\rm O}$ values than indicated
by most Seyferts.  This could be a result of extinction. Seyferts
showing X-ray spectra absorbed by a column greater than $10^{22}$
cm$^{-2}$ (NGC 2639, 4151, 4258, 4388, 4395, 5252, and 5674) are shown
as open circles in Fig. 2. At least four of them have larger $R_{\rm
O}$ than indicated by the correlation. The correlation between $\log
R_{\rm O}$ and $\log R_{\rm X}$ for the less absorbed Seyferts can be
described as $\log R_{\rm O}$ = 0.88 $\log R_{\rm X}$ + 5.0.
According to this relation, the boundary between radio loud and radio
quiet object ($\log R_{\rm O}$ = 1) corresponds to $\log R_{\rm X} =
-4.5$.

The PG quasars show systematically lower $R_{\rm O}$ values than those
of Seyferts at a given $\log R_{\rm X}$.  For the former objects,
$\log R_{\rm O}=1$ corresponds to $\log R_{\rm X}=-3.5$.  This
apparently reflects a luminosity dependence of the shape of the SED:
luminous objects have steeper optical-X-ray slopes $\alpha _{\rm ox} =
1.4-1.7$ ($S\propto\nu^{-\alpha}$), where $\alpha _{\rm ox}$ is often
measured as the spectral index between 2200 A and 2 keV, while less
luminous AGNs have $\alpha _{\rm ox} = 1.0-1.2$ (Ho 1999).  This is
related to the fact that luminous objects show a more prominent ``big
blue bump'' in their spectra.  Figure 8 of Ho (1999) demonstrates that
low-luminosity objects are typically 1--1.5 orders of magnitude
fainter in the optical band than luminous quasars for an given X-ray
luminosity.

The definition of radio loudness using the hard X-ray flux ($R_{\rm
X}$) appears to be more robust because X-rays are less affected by
both extinction at optical wavelengths and the detailed shape of the
blue bump. Further, measurements of nuclear X-ray
fluxes are much easier than measurements of nuclear optical fluxes,
since in the latter case the nuclear light must be separated from the
surrounding starlight.

Fig. 3 shows the X-ray luminosity dependence of $R_{\rm X}$. In this
plot, the LLAGN sample discussed in the present paper is shown in
addition to the Seyfert and PG samples used above. This is an ``X-ray
version'' of the $\log R_{\rm O}$-$M_{B}^{\rm nuc}$ plot (Fig. 4 in Ho
\& Peng 2001). Our plot shows that a large fraction of
LLAGNs ({\LX}$<10^{42}$ {\eps}) are radio loud. This is a confirmation
of Ho \& Peng's (2001) finding. Since radio emission in LLAGNs is likely
to be dominated by emission from jets (Nagar et al. 2001; Ulvestad \&
Ho 2001), these results suggest that, in LLAGN, the fraction of the
accretion energy that powers a jet, as opposed to electromagnetic
radiation, is larger than in more luminous Seyfert galaxies and
quasars.

\begin{figure}
\centerline{\psfig{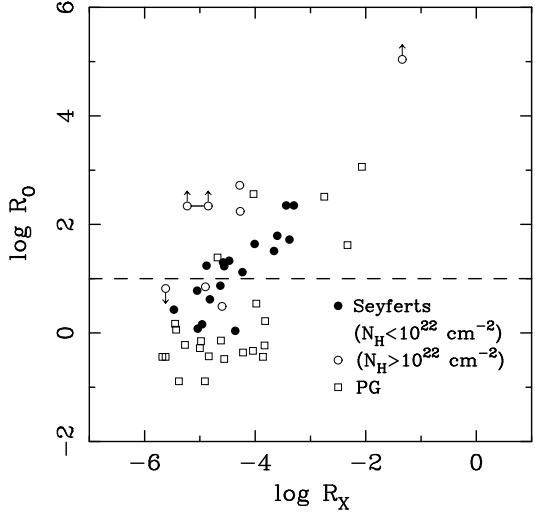}}
\caption[]{
Comparison between $R_{\rm O}=L_{\nu}$(5 GHz)/$L_{\nu}$(B) and 
$R_{\rm X}=\nu L_{\nu}$(5 GHz)/{\LX} for Seyferts and PG quasars. 
The conventional boundary between ``radio loud'' and 
``radio quiet'' objects ($\log R_{\rm O}=1$) is shown as a horizontal 
dashed line.
}
\end{figure}

\begin{figure}
\centerline{\psfig{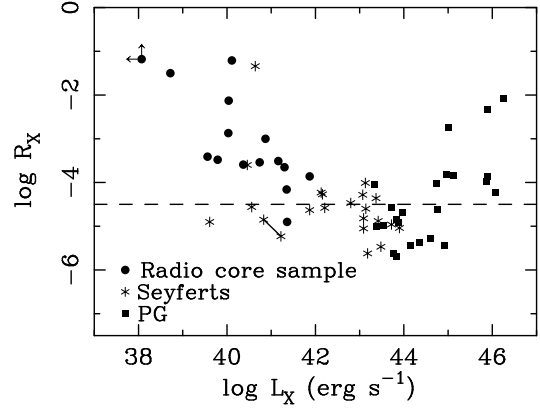}}
\caption[]{
X-ray luminosity dependence of $R_{\rm X}=\nu L_{\nu}$(5 GHz)/{\LX} for the 
present LLAGN sample, Seyfert galaxies, and PG quasars. 
The boundary between ``radio loud'' and 
``radio quiet'' objects ($\log R_{\rm x}=-4.5$) is shown as a horizontal 
dashed line. }
\end{figure}

\begin{acknowledgements}
Y.T. is supported by the Japan Society for a Promotion of Science
Postdoctoral Fellowship for Young Scientists.  This research was
supported by NASA through grants NAG81027 and NAG81755 to the
University of Maryland.
\end{acknowledgements}


\vspace{2cm}

\begin{thebibliography}{}

\bibitem{} Falcke, H., Nagar, N.~M., Wilson, A.~S., \& Ulvestad, J.,~S. 2000, ApJ, 542, 197

\bibitem{} Finoguenov, A. \& Jones, C. 2001, ApJ, 547, L107	

\bibitem{} Ho, L.~C. 1999, ApJ, 516, 672

\bibitem{} Ho, L.~C., et al. 2001, ApJ, 549, L51

\bibitem{} Ho, L.~C., Filippenko, A.~V., \& Sargent, W.~L.~W. 1997a, ApJS, 112, 315

\bibitem{} Ho, L.~C., Filippenko, A.~V., Sargent, W.~L.~W., \& Peng, C.~Y. 1997b, ApJS, 112, 391

\bibitem{} Ho, L.~C., \& Peng, C.~Y. 2001, ApJ, 555, 650

\bibitem{} Nagar, N.~M., Falcke, H., Wilson, A.~S., \& Ho, L.~C. 2000, ApJ, 542, 186

\bibitem{} Nagar, N.~M., Falcke, H., Wilson, A.~S., \& Ulvestad, J.~S. 2002, A\&A, submitted

\bibitem{} Nagar, N.~M., Wilson, A.~S., \& Falcke, H. 2001, ApJ, 559, L87

\bibitem{} Terashima, Y., Ho, L.~C., Iyomoto, N., \& Ptak, A.~F. 2002a, ApJ, in preparation

\bibitem{} Terashima, Y., Ho, L~.C., \& Ptak, A.~F. 2000a, ApJ, 539, 161 

\bibitem{} Terashima, Y., Ho, L~.C., Ptak, A.~F., et al. 2000b, ApJ, 533, 729      

\bibitem{} Terashima, Y., Iyomoto, N., Ho, L.~C., \& Ptak, A.~F. 2002b, ApJS, 139, 1

\bibitem{} Terashima, Y. \& Wilson, A.~S. 2002c, ApJ, submitted

\bibitem{} Turner, M.~J., et al. 2001, A\&A, 365, L110

\bibitem{} Turner, T.~J., George, I.~M., Nandra, K., \& Mushotzky, R.~M. 1997, ApJS, 113, 23

\bibitem{} Ulvestad, J.~S., \& Ho, L.~C. 2001, ApJ, 562, L133

\bibitem{} Wilson, A.~S. \& Yang, Y 2002, ApJ, 568, 133

\end{thebibliography}
\end{document}